\pdfoutput=1
\documentclass[twocolumn,english,aps,superscriptaddress,prb]{revtex4-1}

\usepackage{babel}
\usepackage{amsmath, amsfonts, amssymb, mathrsfs, mathtools, stmaryrd}
\usepackage{graphicx, color}
\usepackage{xargs}                      
\usepackage[pdftex,dvipsnames]{xcolor}  
\usepackage{hyperref}
\hypersetup{
   colorlinks,
   citecolor=blue,
   filecolor=blue,
   linkcolor=blue,
   urlcolor=blue,
   breaklinks=true
}

\begin{document}

\title{Possible restoration of particle-hole symmetry in the 5/2 Quantized Hall State at small magnetic field}

 \author{Lo\"ic Herviou}
 \affiliation{Institute of Physics, Ecole Polytechnique F\'ed\'erale de Lausanne (EPFL), CH-1015 Lausanne, Switzerland}
 \author{Fr\'ed\'eric Mila}
 \affiliation{Institute of Physics, Ecole Polytechnique F\'ed\'erale de Lausanne (EPFL), CH-1015 Lausanne, Switzerland}

\date{\today}
\begin{abstract} 
Motivated by the experimental observation of a quantized 5/2 thermal conductance at filling $\nu=5/2$, a result incompatible with both the Pfaffian and the Antipfaffian states, we have pushed the expansion of the effective Hamiltonian of the $5/2$ quantized Hall state to third-order in the parameter $\kappa=E_c/\hbar \omega_c\propto 1/\sqrt{B}$ controlling the Landau level mixing , where $E_c$ is the Coulomb energy and $\omega_c$ the cyclotron frequency. Exact diagonalizations of this effective Hamiltonian show that the difference in overlap with the Pfaffian and the AntiPfaffian induced at second-order is reduced by third-order corrections and disappears around $\kappa=0.4$, suggesting that these states are much closer in energy at smaller magnetic field than previously anticipated. 
Furthermore, we show that in this range of $\kappa$ the finite-size spectrum is typical of a quantum phase transition, with a strong reduction of the energy gap and with level crossings between excited states.
These results point to the possibility of a quantum phase transition at smaller magnetic field into a phase with an emergent particle-hole symmetry that would explain the measured $5/2$ thermal conductance of the $5/2$ quantized Hall state. 
\end{abstract}

\maketitle


\section{Introduction}

Following up on Laughlin's pioneering explanation of the $1/3$ and $1/5$ plateaus\citep{Laughlin1983} in terms of wave-functions, the theory of the Fractional Quantum Hall Effect (FQHE) has relied to a large extent on variational wave-functions, with considerable success thanks in particular to Jain's composite fermion theory\citep{Haldane1983, Halperin1984, Jain1989, BookHalperinJain}.
It was completed by its extension to the Pfaffian (Pf) state by Moore and Read\citep{Moore1991, Read1992}, and its particle-hole conjugate, the AntiPfaffian\citep{Levin2007, Lee2007} (APf) state in order to explain the plateau observed at filling $5/2$\citep{Willet1987, Xia2004}.
This approach has explained many plateaus and has led to the highly nontrivial prediction that the system is gapless at filling $1/2$\citep{Halperin1993, Willet1993, Scarola2000}. 
It seemed that all states observed in the FQHE could be explained in terms of variational wave-functions. 
The alternative approach in terms of an effective Hamiltonian to describe the degeneracy lifting by Coulomb repulsion in a partially filled Landau level has nevertheless proven to be extremely useful. 
The variational wave-functions were shown to be the ground states of parent Hamiltonians that are truncated versions of the effective model\citep{Haldane1983, Greiter1991, Lee2015, Chen2017, Bandyopadhyay2020}.
Effective Hamiltonians have also played a crucial role in discussing the competition between the Pf and the APf states for the 5/2 quantized Hall state, with the conclusion that, to second order in $\kappa=E_c/\hbar \omega_c$, where $E_c$ is the Coulomb energy and $\omega_c\propto B$ the cyclotron frequency, the APf state is favored when the effect of the empty Landau levels are taken into account\citep{Wojs2010, Rezayi2011, Peterson2013, Simon2013, Sodemann2013, Pakrouski2015, Zaletel2015, Rezayi2017}. 
It thus came as a big surprise when the quantized thermal conductance was found to be equal to $5/2$ in the $5/2$ plateau\citep{Banerjee2018, Dutta2021, Dutta2022}, a value in contradiction with both the APf ($3/2$) and the Pf ($7/2$) but consistent with particle-hole symmetry. 
This restored symmetry would be consistent with the Particle-Hole Pfaffian (PHPf)\citep{Jolicoeur2007, Son2015}, another candidate for the $5/2$ plateau, but this state is usually believed to be gapless and energetically unfavored\citep{Balram2018, Mishmash2018, Yutushui2020, Rezayi2021} except in a few field-theoretical works\citep{Antonic2018, Djurdevic2019}.
Explanations in terms of domains or of lack of equilibration have  been put forward\citep{Zucker2016, Simon2018, Wang2018, Mross2018, Lian2018, Feldman2018, Simon2018-2, Ma2019, Park2020, Fulga2020, Simon2020, Simon2020-2, Asasi2020, Ma2020}, but as of today there is no consensus on the resolution of this discrepancy.\\

In view of the impressive corpus of theory, one may wonder if there is still room for the identification of a particle-hole symmetric ground state that would have been missed so far. 
Our results point towards such an unlikely conclusion.
The starting point of our approach is the observation that the effective model is an expansion in $\kappa \propto 1/\sqrt{B}$, hence a high-field expansion. 
In experimental conditions, the field is around $4$ T, and $\kappa \simeq 1.38$\citep{Xia2004}. 
This large value raises a natural question: Is $\kappa$ too large in experiments for the second-order expansion to be justified?
The only way to answer that question is to push the expansion to higher-order in $\kappa$, something that has not been attempted so far.\\

In the present work, we have pushed this expansion to the next order in $\kappa$. 
Although this simply relies on third-order perturbation theory in the Coulomb repulsion, this turned out to be a rather formidable task that could only be carried out with the help of computer-aided formal calculations. 
As we shall see, the third order changes the physics qualitatively already around $\kappa \simeq 0.3-0.5$, showing that relying on second-order perturbation theory is definitely not justified for $\kappa \simeq 1.38$. 
The most remarkable effect is that the lifting of the degeneracy in favour of the APf is counter-acted by the third-order term, and that the degeneracy is restored at $\kappa \simeq 0.4$. 
Moreover, and maybe more importantly, the excitation spectrum of finite-size systems has all the characteristics of a quantum phase transition which, in view of the apparent restoration of particle-hole symmetry between the Pf and the APf, might lead to a phase with an emergent particle-hole symmetry. 
Let's already emphasize however that, if there is indeed a quantum phase transition, the physics beyond the phase transition cannot be reached by perturbation theory, and any attempt at discussing it on the basis of a truncated perturbative Hamiltonian, as done previously for the second-order model, is prone to fail. 
Non-perturbative approaches will have to be employed to study that problem.

The paper is organized as follows. In Section II, we explain the main ideas of the algorithm we have used to derive the third-order Hamiltonian. In Section III-A, we compare the results we have obtained at second-order with previous results, and in Section III-B we present and discuss the central results of this paper obtained at third-order. Section IV is devoted to two related models that help assessing the validity of our approach, a model that only includes the first two Landau levels (IV-A), and a model that assumes full polarization of the lowest Landau level (IV-B). Finally, the implications for the 5/2 quantum Hall state are discussed in Section V. Details about all aspects of this study can be found in the Appendices.

\section{Derivation of the effective Hamiltonian}
We consider two-dimensional electrons on a square torus in a normal magnetic field and in the presence of Coulomb interaction.
Up to a constant, the Hamiltonian can be formulated in a second-quantization formalism as
\begin{align}
\mathcal{H}_\mathrm{exact} &= \mathcal{H}_0 + \mathcal{H}_1 \label{eq:Hamexact}\\
\mathcal{H}_0 &= \hbar \omega_c \sum\limits_{l} l  N_l - gB \sum\limits_{\sigma}  \sigma N_\sigma\\
\mathcal{H}_1 &= E_c \sum\limits_{\vec{m}, \vec{n}, \vec{l}, \vec{\sigma}} A_{\vec{m}, \vec{n}}^{\vec{l}, \vec{\sigma}} c^\dagger_{m_1, l_{m_1}, \sigma_{m_1}}  c^\dagger_{m_2, l_{m_2}, \sigma_{m_2}} \nonumber \\
	&\qquad \qquad \qquad \qquad \times c_{n_2, l_{n_2}, \sigma_{n_2}} c_{n_1, l_{n_1}, \sigma_{n_1}}.
\end{align}
$\omega_c$ is the cyclotron frequency, $E_c = \frac{e^2}{\varepsilon l_B}$ the Coulomb energy, $l_B$ the magnetic length, $B$ the magnetic field and $g$ the electronic magnetic moment.
In the rest of the article, we set both $l_B$ and $E_c$ to $1$ for simplicity, and generally neglect the Zeeman splitting given its magnitude.
$l \in \mathbb{N}$ denotes the Landau level, $\sigma = \pm 1$ its spin flavour, $N_l$ the number of electrons in a given Landau level and $N_\sigma$ the number of electrons with spin $\sigma$.
We also denote by $L$ the total number of Landau levels we consider (the spin degeneracy is not included in the counting).
We will show results with up to $L = 11$, although our results depend very little on $L$ as soon as $L\geq 3$, i.e. when we take into account the influence of the empty Landau levels.
Finally, the operator $c^\dagger_{m, l, \sigma}$ creates an electron in the $m$th orbital of the $l$th Landau level with spin $\sigma$.
On a torus, $m \in [0, N_\phi-1]$ with $N_\phi$ the number of elementary magnetic fluxes through the torus.
The interaction coefficient $A_{\vec{m}, \vec{n}}^{\vec{l}, \vec{\sigma}}$ can be straightforwardly obtained for any translation-invariant interaction as detailed in App.~\ref{app:Algo}. 
For a Coulomb-like interaction (central and spin-diagonal), symmetry enforces $m_1 + m_2 = n_1 + n_2 \, [ N_\phi ] $ and $\sigma_{m_1} + \sigma_{m_2} = \sigma_{n_1} + \sigma_{n_2}$.\\

In the rest of this article, we focus on the physics of the $5/2$ filled Landau levels.
In a strong magnetic field, the splitting of the Landau levels dominates, and it appears reasonnable to project the Hamiltonian on its low-energy sector (depending on the filling).
Despite the weak Zeeman effect, numerical simulations seem to indicate that the half-filled Landau level is spin polarized.
We therefore introduce $P_0$, the projector on the subspace where the $0$th Landau level is fully occupied for both spin flavors, and where the $1$st Landau level with spin $+1$ is half-filled.
We project the Hamiltonian $\mathcal{H}_\mathrm{exact}$ on this subspace and define:
\begin{equation}
H_0 = P_0 \mathcal{H}_0 P_0  = E_0 \mathrm{Id} \text{ with } E_0 = (\hbar \omega_c - g B)  \frac{N_\phi}{2}
\end{equation}
\begin{equation}
H_1 = P_0 \mathcal{H}_1 P_0 = E_c \sum\limits_{\vec{m}, \vec{n}} A_{\vec{m}, \vec{n}}^{\vec{l}, \vec{\sigma}} c^\dagger_{m_1}  c^\dagger_{m_2} c_{n_2} c_{n_1} + E_C^{0,1},
\end{equation}
with $c_{n} = c_{n, 1, +1}$ and $E_C^{0,1}$ the static energy due to the presence of the filled Landau levels.
Note that this energy constant plays a crucial role in the third-order expansion and cannot be neglected.
This projected Hamiltonian with no Landau-level mixing has been extensively studied\citep{Yoshioka1984, Greiter1991, Morf1998, Park1998, Rezayi2000, Wojs2001}.
On the square torus, it admits six quasi-degenerate ground states in six different translation sectors corresponding to the six-fold topological degeneracy of the Pf or APf state.
Unless specified, we show results in the $(\pi, 0)$ sector --- our results are largely independent of this choice. 
This Hamiltonian  is particle-hole symmetric at half-filling.
Due to this symmetry, it cannot discriminate between the Pf and APf phases.
The ground state in a given sector is unique and has equal overlap with both Pf and APf states. \\

In order to reach the experimentally relevant regime, we compute perturbatively the effect of the presence of the empty and occupied Landau levels using $\kappa = \frac{e^2}{\varepsilon \hbar l_B \omega_c}$ as a small parameter.
Following Rezayi\citep{Rezayi2017}, who performed a calculation to second order, we compute directly the effective Hamiltonian without attempting to project onto pseudo-potentials.
The development in pseudo-potentials is not convenient on a finite torus due to the periodicity, and the complexity of the higher-body terms rises quickly.
The second-order and third-order terms of the degenerate perturbative expansion are given by
\begin{align}
H_2 &= - P_0 \mathcal{H}_1 \mathcal{G}_0  \mathcal{H}_1 P_0 \nonumber\\
H_3 &= P_0 \mathcal{H}_1 \mathcal{G}_0  \mathcal{H}_1 \mathcal{G}_0  \mathcal{H}_1 P_0 - \frac{1}{2} \{H_1, P_0 \mathcal{H}_1 \mathcal{G}^2_0  \mathcal{H}_1 P_0  \} \label{eq:3rdOrderSW}
\end{align}
where 
\begin{equation}
\mathcal{G}_0 = \frac{\mathrm{Id} - P_0}{\mathcal{H}_0 - E_0}.
\end{equation}
$H_2$ includes two- and three-body terms as discussed in previous works.
$H_3$ also includes an additional four-body term.
The five-body terms generated by $P_0 \mathcal{H}_1 \mathcal{G}_0  \mathcal{H}_1 \mathcal{G}_0  \mathcal{H}_1 P_0$ are exactly cancelled by the anticommutator.\\

We perform the numerical computation of the effective Hamiltonians directly at the operator level.
The details of our algorithm can be found in App.~\ref{app:Algo}.
As a quick summary, our computational process consists of four parts:
\begin{enumerate}
\item Computation of the effective interaction in the Landau basis.
\item Derivation of all the Feynman diagrams corresponding to $\mathcal{H}_2$ and $\mathcal{H}_3$.
\item Exact summation of all processes corresponding to a given diagram.
\item Computation of the effective many-body Hamiltonian and diagonalization.
\end{enumerate}
The latter two steps are the most computationaly expensive.
The complexity of the third step scales as $O(\max(  5! L N_\phi^7 , (2L)^4 N_\phi^4 ))$, depending on the diagrams considered.
The fourth step has the standard exponential complexity of exact diagonalization, but with an additional difficulty: the effective Hamiltonian consists of a sum of several millions of $n$-body operators.
To give a concrete illustration, for $N_\phi = 28$, although the symmetry-resolved Hilbert space is only of dimension $\approx 10^5$, it includes $\approx 1.5 \times 10^7$ operators ($\approx 10^6$ if we take into account translation invariance).
Even if we were to discard coefficients below $10^{-6}$, we would need to apply several millions of operators to each basis element.
Consequently, it is not surprising that the effective Hamiltonians themselves are also very dense (approximately a quarter of the matrix elements are non-zero for $N_\phi = 28$).
This density limits the practically achievable sizes significantly: both the memory cost to store the matrix and the cost of applying it to a state become quickly prohibitive. 

\section{Perturbative expansion}

\subsection{Second-order expansion}
We start by a brief discussion of the second-order expansion of $\mathcal{H}_\mathrm{exact}$ as a benchmark of our approach.
This computation has been previously done on the sphere \citep{Wojs2010, Rezayi2011, Peterson2013, Simon2013, Pakrouski2015, Zaletel2015, Rezayi2017, Das2022} and on the hexagonal torus \citep{Pakrouski2015, Rezayi2017} and we verify that we qualitatively and quantitatively recover known results.\\

\begin{figure}[t!]
\centering\includegraphics[width=0.48\textwidth]{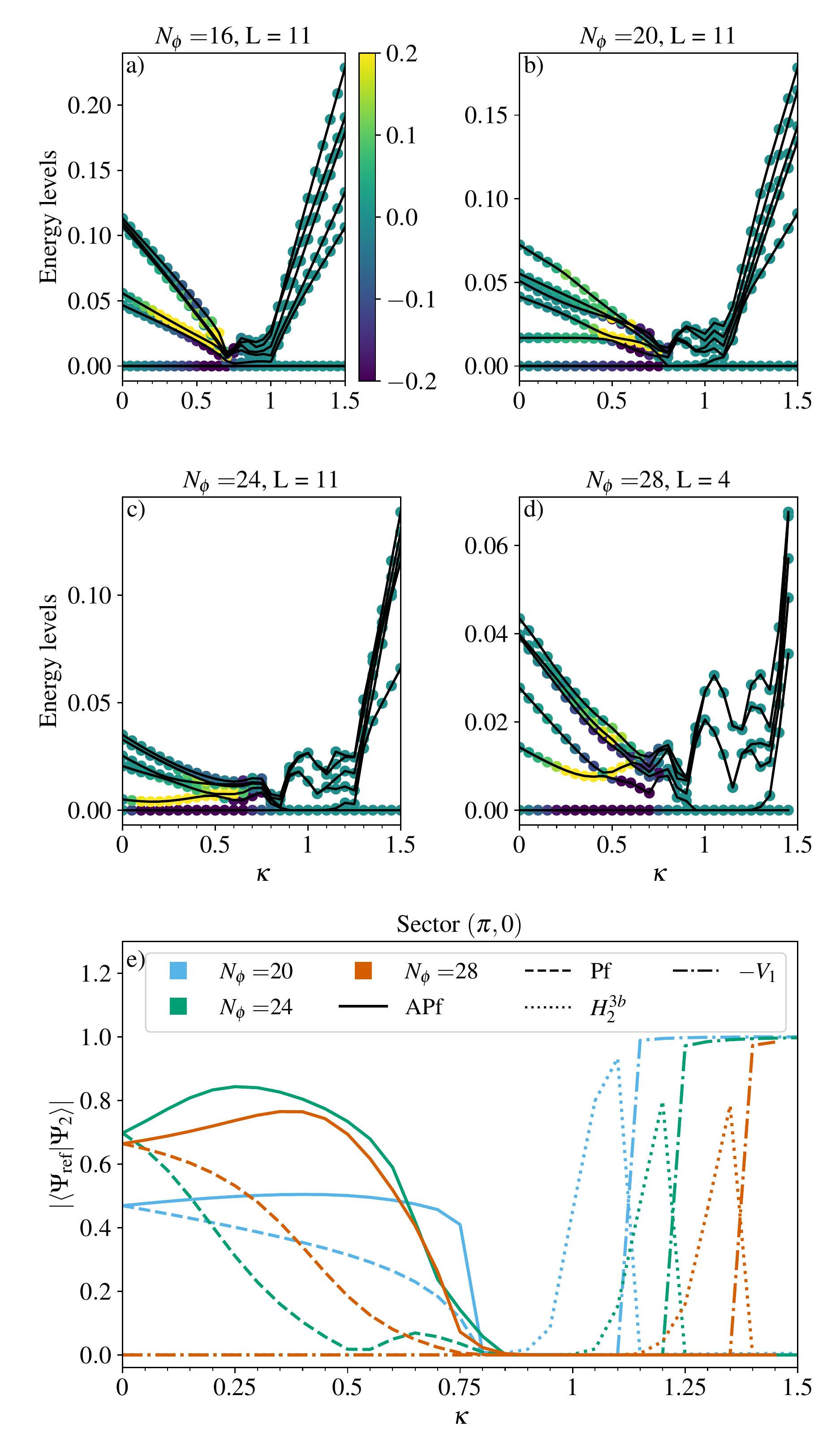}
\caption{(a-d) Low-energy spectrum in the sector $(\pi, 0)$ at second order.
The color code is given in Eq.~\eqref{eq:ColorCode}: positive (resp. negative) numbers mark that the Pf (resp. APf) state is favored.
(e) Overlap between the ground state and several reference states for different system sizes. 
Here the overlap is not corrected as in Eq.~\eqref{eq:ColorCode} for simplicity.
Below $\kappa \approx 0.8$, the APf is favored. We then observe a collapse of all low-energy levels. %
After a small transitional regime, two consecutive phases open. 
The large $\kappa\geq 1.4$ regime is strongly gapped in the translation invariant sector.}
\label{fig:spectrum_2ndorder}
\end{figure}

Concretely, we work with the Hamiltonian 
\begin{equation}
H^{(2)} = H_1 + \kappa H_2
\end{equation}
and compute its ground state.
In Fig.~\ref{fig:spectrum_2ndorder}\hyperlink{fig:spectrum_2ndorder}{(a-d)}, we show the low-energy spectrum of $H^{(2)}$.
The color represents the difference in overlaps between the Pf and APf.
More precisely, the color is a measure of
\begin{equation}
\vert \langle \Psi_\mathrm{Pf} \vert \Psi_\mathrm{ED} \rangle_c \vert - \vert \langle \Psi_\mathrm{APf} \vert \Psi_\mathrm{ED} \rangle_c \vert\label{eq:ColorCode}  
\end{equation}
\begin{equation}
\text{with } \begin{pmatrix}
\langle \Psi_\mathrm{Pf} \vert \Psi_\mathrm{ED} \rangle_c \\ \langle \Psi_\mathrm{APf} \vert \Psi_\mathrm{ED} \rangle_c
\end{pmatrix} = M^{-1} \begin{pmatrix}
\langle \Psi_\mathrm{Pf} \vert \Psi_\mathrm{ED} \rangle \\
\langle \Psi_\mathrm{APf} \vert \Psi_\mathrm{ED} \rangle \label{eq:defCorrOver}
\end{pmatrix}
\end{equation}
\begin{equation}
\text{ and } M = \begin{pmatrix}
\langle \Psi_\mathrm{Pf} \vert \Psi_\mathrm{Pf} \rangle & \langle \Psi_\mathrm{Pf} \vert \Psi_\mathrm{APf} \rangle\\
\langle \Psi_\mathrm{APf} \vert \Psi_\mathrm{Pf} \rangle & \langle \Psi_\mathrm{APf} \vert \Psi_\mathrm{APf} \rangle \label{eq:defOverMat}
\end{pmatrix}.
\end{equation}
In Fig.~\ref{fig:spectrum_2ndorder}\hyperlink{fig:spectrum_2ndorder}{e}, we show the overlap of the corresponding ground state with several reference states.
At small $\kappa$, for $L \geq 3$, the second-order expansion favors the APf state (for $L=2$, the Pf is actually favored).
At $\kappa \approx 0.8$, the gap closes and a large number of low-excited states collapse onto the ground state.
After a transitory regime, a small gap opens.
Its ground state is adiabatically connected to the ground state of $H_2^\mathrm{3b}$, the three-body contribution of $H_2$.
A second transition then occurs towards the ground state of $H_2$  (which is approximately also the ground state of $H_2^\mathrm{2b}$). \\

It is important to note the following.
Firstly, the collapse of the energy levels we observe at $\kappa \approx 0.8$ is qualitatively different from the typical second-order phase transitions in finite systems.
Instead of well-defined minimum that decreases with the system size, we observe several crossings or anti-crossings in the ground state over a range of $\kappa$.
Secondly, the two large $\kappa$ phases correspond to a limit where the perturbation theory dominates and are unphysical.
The ground state of $H_2^\mathrm{2b}$ has nearly perfect overlap with the maximally excited state of the first two-body Haldane pseudopotential $V_1$.
In other words, it is the ground state of a globally attractive potential.
Additionally, it becomes $2N_\phi$ degenerate in the limit $\kappa \gg 1$ (in contrast, the topological degeneracy of both the Pf and the APf is equal to $6$) with ground states in each sector of the form $(\pi, . )$ and $(., \pi)$.
It is particle-hole symmetric, and, due to the periodicity of the torus, it is not a simple packed state.
The recent results of Ref.~\onlinecite{Das2022} could potentially be explained by a state similar to either of the two large $\kappa$ phases.

\subsection{Third-order perturbation theory}
We now discuss our results for the third-order Hamiltonian
\begin{equation}
H^{(3)} = H_1 + \kappa H_2 + \kappa^2 H_3.
\end{equation}
We first investigate the effect of $H^{(3)}$ directly on the reference Pf and APf wavefunctions, before turning to the study of its ground state.\\

The Pf and APf have a large overlap with the low-energy eigenstates of the Coulomb Hamiltonian $H_1$ and are therefore a good first ansatz to qualitatively study the model.
If there is an emergent particle-hole symmetry at low energies, they should become degenerate.
We therefore investigate the average energy in both states and define the energies per orbital
\begin{equation}
E_\mathrm{ref} = \frac{1}{N_\phi} \langle \Psi_\mathrm{ref} \vert H^{(3)} \vert \Psi_\mathrm{ref} \rangle. \label{eq:effEner}
\end{equation}
For $L=11$, $N_\phi = 24$ in the $(\pi, 0)$ subsector, we obtain
\begin{align}
E_\mathrm{Pf} &= -1.3540 -0.3083 \kappa + 0.1015\kappa^2 ,\\
E_\mathrm{APf} &= -1.3540 -0.3090 \kappa + 0.1033\kappa^2 
\end{align}
Due to the particle-hole symmetry, $H_1$ does not discriminate between these two states, and, in agreement with Rezayi\citep{Rezayi2017}, $H_2$ favors the APf.
By contrast, $H_3$ favors the Pf.
More precisely, its three-body contribution $H_3^\mathrm{3b}$ favors the Pf while the four-body contribution $H_3^\mathrm{4b}$ still advantages the APf.
For $\kappa_c \approx 0.42$, the Pf and the APf are degenerate.
We therefore expect a restoration of the particle-hole symmetry in the low-energy subspace near this critical $\kappa$.
Note that, due to finite-size effects, they can have a significant overlap with each other, but are orthogonal in the thermodynamic limit.
A more complete analysis of the Pf and APf subspace, along with data for different $L$ and $N_\phi$, can be found in App.~\ref{app:effHam}.
Our conclusions are unaffected.\\

\begin{figure}[t!]
\centering\includegraphics[width=0.45\textwidth]{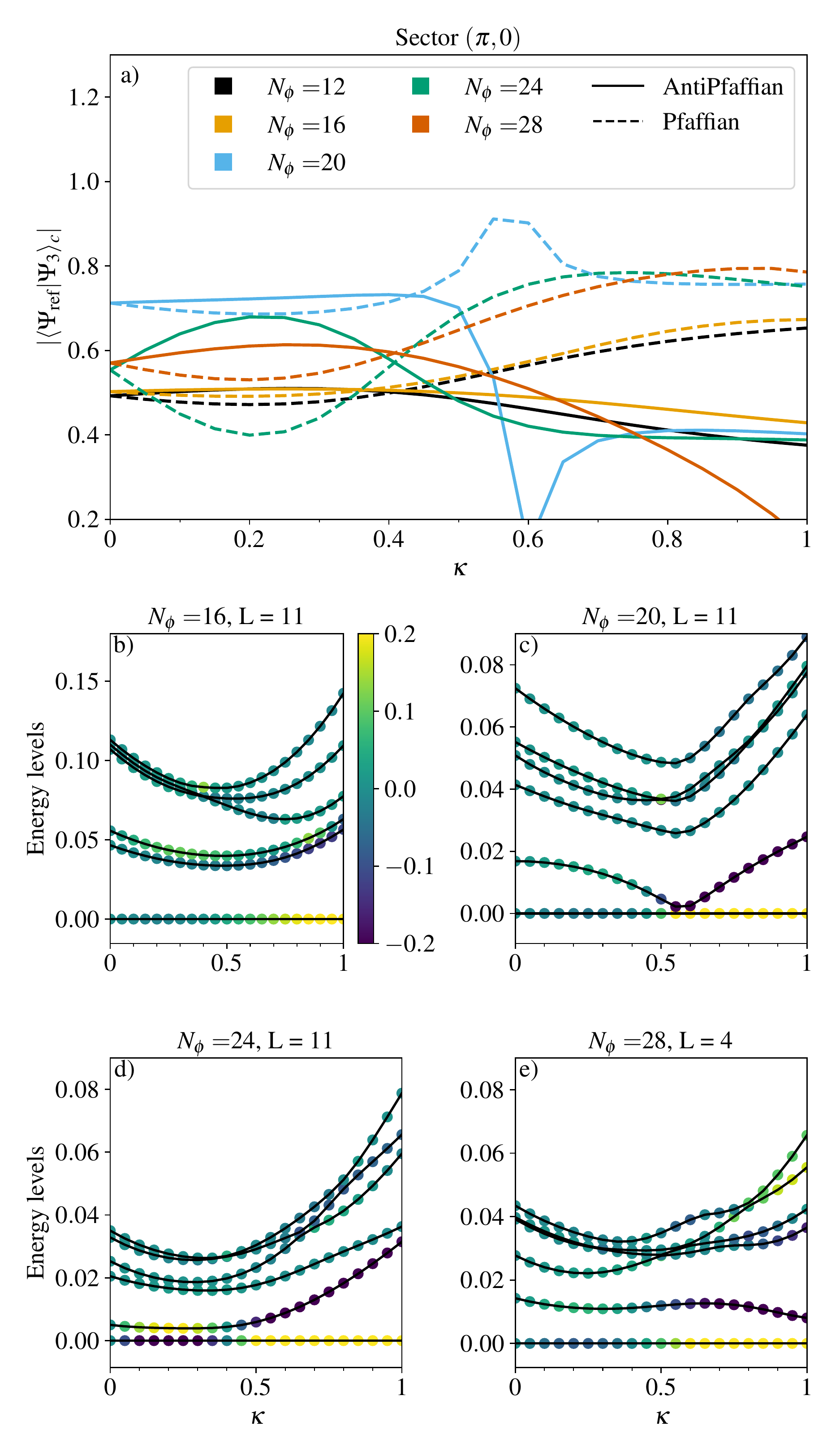}
\caption{(a) Overlaps of the third-order ground state with the Pf and APf. All overlaps are corrected following Eq.~\eqref{eq:ColorCode}. (b-e) Low-energy spectra. Below $\kappa_c \approx 0.4$, the APf is favoured, while for larger $\kappa$, it is the Pf. We observe a series of energy level crossings strongly reminiscent of a second-order phase transition.  
}
\label{fig:overlap}
\end{figure}

\begin{figure}
\centering\includegraphics[width= 0.4\textwidth]{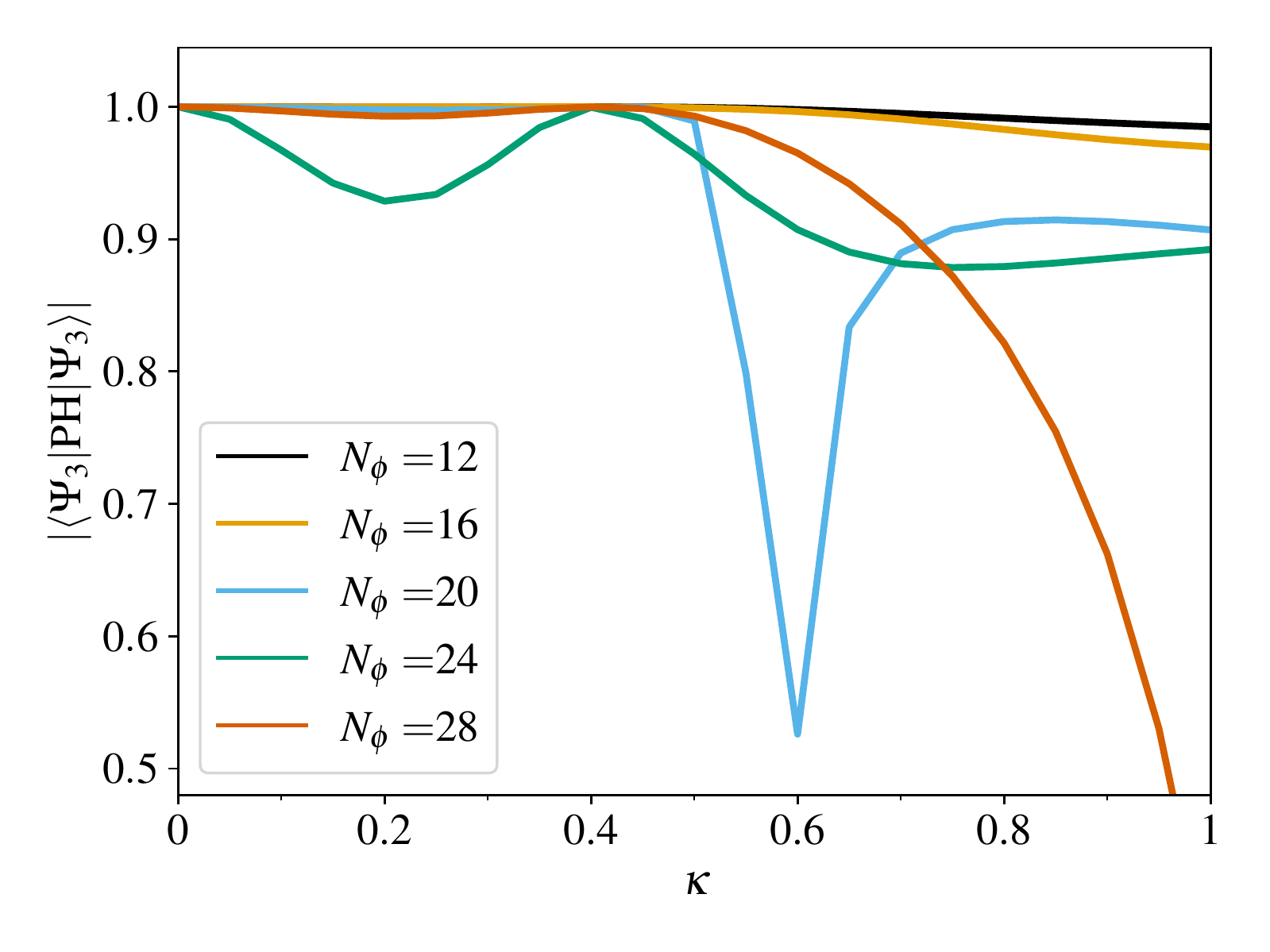}
\caption{Overlaps between the third-order ground state and its particle-hole symmetric partner. For all system sizes, the particle-hole symmetry is restored around $\kappa = 0.4$. Note that for most system sizes, it is only weakly broken at low $\kappa$.}
\label{fig:PHsym}
\end{figure}

We now turn to the study of the ground state of $H^{(3)}$ to confirm this naive approach.
In Fig.~\ref{fig:overlap}\hyperlink{fig:overlap}{a}, we show the overlaps of the third-order ground state with the Pf and the APf, corrected with Eq.~\eqref{eq:ColorCode}.
While the APf is still favored at small $\kappa$, the amplitude of the difference in overlaps is significantly reduced.
For $\kappa \geqslant 0.2$, the difference starts decreasing and at $\kappa_c \approx 0.4$, the ground state has equal overlap with the Pf and the APf.
$\kappa_c$ depends very little on the system size and the ground state becomes particle-hole symmetric as shown in Fig.~\ref{fig:PHsym}.
The low-energy spectrum depicted in Fig.~\ref{fig:overlap}\hyperlink{fig:overlap}{b-e} presents numerous level crossings and a curvature coherent with the finite-size spectrum of a second-order phase transition.
Finally, for larger $\kappa$, the ground state has a large overlap with the Pf state.
All these results are in agreement with the previous naive analysis of Eq.~\eqref{eq:effEner}.
They suggest a possible restoration of the particle-hole symmetry at intermediate $\kappa$, which could explain the experimental thermal conductance.
In the following Section, we discuss the validity of this hypothesis.\\

\begin{figure}
\centering\includegraphics[width = 0.85\linewidth]{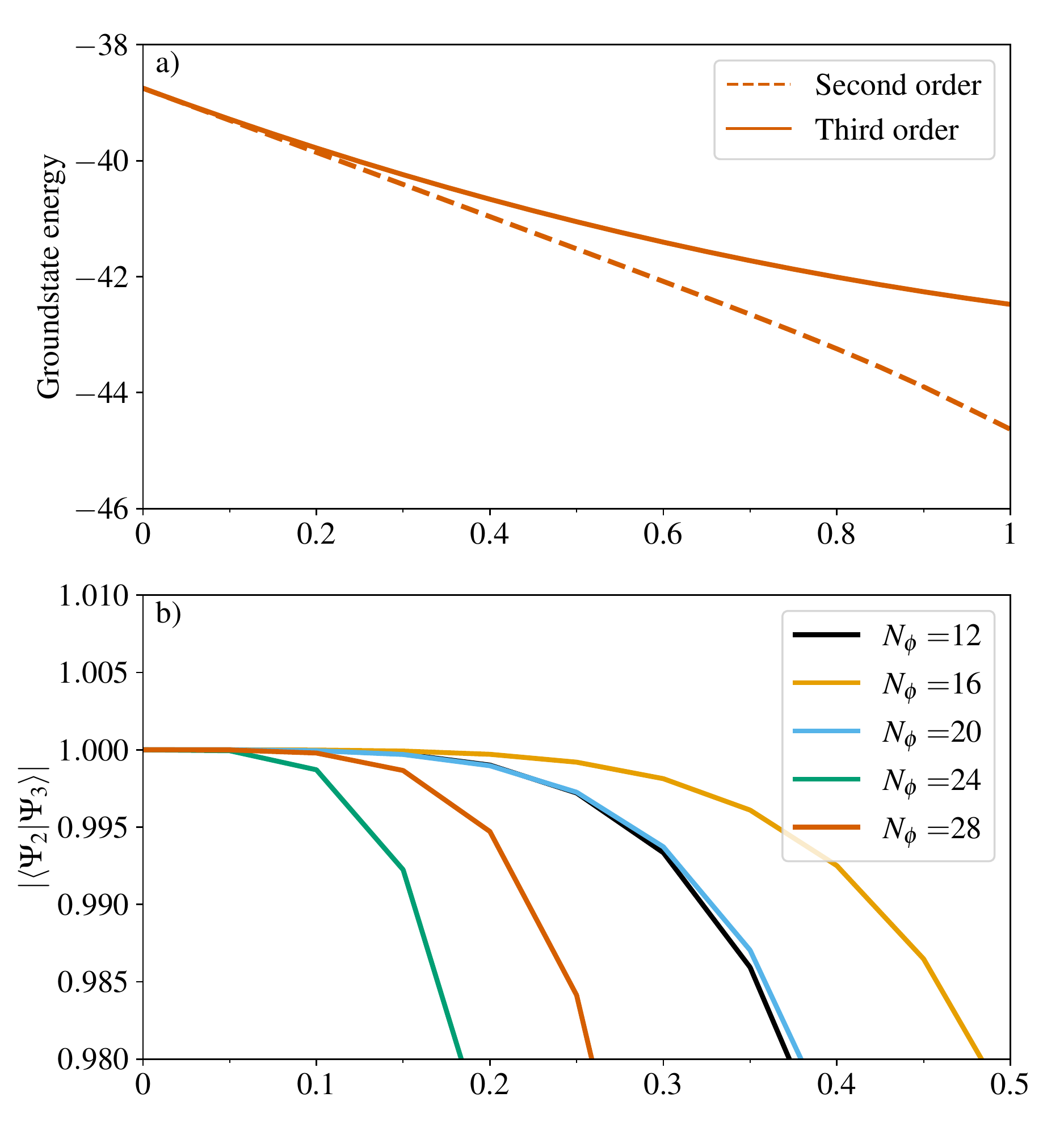}
\caption{a) Ground state energy for second- and third-order expansion at $N_\phi = 28$ and $L = 3$. Visible differences are already present at $\kappa \approx 0.2$. b) Overlap between the ground states of $H^{(2)}$ and $H^{(3)}$ at small $\kappa$. The expansions agree with an overlap $\geq 0.99$ up to $\kappa \approx 0.15$ only.}
\label{fig:comparison23}
\end{figure}

Before that, let us assess the limit of validity of the second-order expansion.
In Fig.~\ref{fig:comparison23}\hyperlink{fig:comparison23}{a}, we compare the ground state energies of $H^{(2)}$ and $H^{(3)}$ for $N_\phi = 28$ and in Fig.~\ref{fig:comparison23}\hyperlink{fig:comparison23}{b} we measure the overlaps between the two ground states for different values of $N_\phi$.
The energies already differ at $\kappa = 0.2$ and the overlaps start falling in the range $0.1 \leq \kappa \leq 0.3$, depending on $N_\phi$.
The second-order expansion therefore cannot be trusted to describe the experimentally relevant regime.

\section{Two related models}

To put our results in perspective, we discuss in this section two related models: (i) The model $L=2$ obtained by keeping only the lowest Landau level for the virtual states; (ii) The fully polarized case where all Landau levels are assumed to be filled only by electrons with a given spin polarization. 
One of the main reasons is that, for both models, we can compare the third-order expansion in $\kappa$ with other approaches, with very encouraging results. 
Interestingly, the physics of these two models turn out to be significantly different from that of the full model, showing that it is crucial to include at least one unoccupied Landau level, and to take into account both spin orientations in the lowest Landau level.

\subsection{The $L=2$ model}\label{sec:3exact}

The main motivation to investigate this model is that, if we keep only the lowest Landau level, we can perform exact diagonalizations and compare the results to a perturbative expansion in $\kappa$. 
The dimension of the Hamiltonian grows as $O(N_\phi^{-2} \binom{LN_\phi}{N_{\sigma = +1}} \binom{LN_\phi}{N_{\sigma = -1}} )$ after taking into account charge, spin and momentum conservation, and translation invariance. As a consequence, the only non-trivial case on which we could do exact diagonalisations is $N_\phi = 8$ fluxes per Landau level and $L = 2$.
Indeed, $N_\phi = 8 $ is the smallest system size where one can see a difference between Pf and APf, and going 
to $L=3$ would mean working with a Hilbert space of dimension $250$ billions (without taking into account translation invariance, but with charge, spin and momentum conservation).

\begin{figure}[t!]
\centering\includegraphics[width=\linewidth]{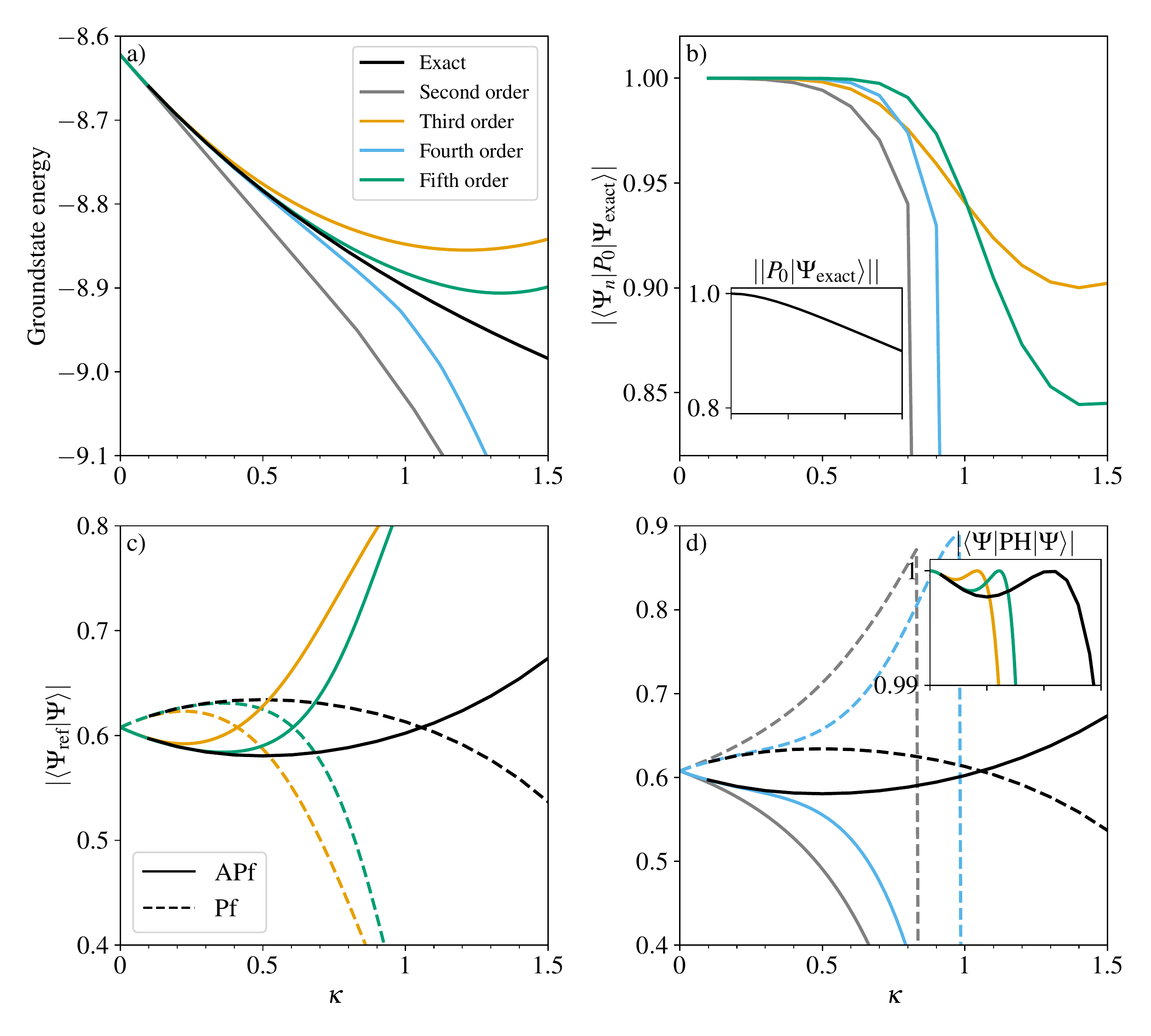}
\caption{Comparison between the exact calculation and several orders of the perturbative expansion for spinful electrons with $N_\phi = 8$ and $L=2$. a) Ground state of the exact Hamiltonian and the first five orders of the expansion. b) Overlaps between the perturbative ground states and the normalized projection of the exact ground state. In the inset, we show the norm of the projected ground state. c-d) Comparison of the overlaps with the Pf and the APf for odd (c) and even (d) orders. We observe the same overlaps crossing for odd-orders as in the exact limit, albeit at much smaller $\kappa$. In the inset, overlaps between the ground states and their particle-hole symmetric partners for the exact and odd-order states.}
\label{fig:exact}
\end{figure}

We show in Fig.~\ref{fig:exact}\hyperlink{fig:exact}{a} the ground state energy for the exact model and the first five orders of the perturbative expansion.
Even for such small systems, we observe a significant deviation of the energy of the second-order correction already at $\kappa \approx 0.2$.
Third-order perturbation extends this range of validity up to $\kappa \approx 0.35$.
The fifth order is only valid up to $\kappa \approx 0.65$.

In Fig.~\ref{fig:exact}\hyperlink{fig:exact}{b}, we show the evolution of the overlaps between the perturbative ground states and the projection of the exact ground state on the spin-polarized manifold. 
The norm of the projection is represented in the inset. 
If one stops at an even order, the overlap with the exact ground state falls dramatically as soon as finite-order effects appear. By contrast, the overlap remains significant if one stops at an odd order. The difference comes from the fact that the last term of the expansion is attractive if it is even, and this is not representative of the exact model at large $\kappa$. So, to discuss the physics beyond the small $\kappa$ range on the basis of a truncated expansion, it is better to stop at an odd order.

More importantly, in Figs.~\ref{fig:exact}\hyperlink{fig:exact}{c-d}, we show the overlap of the different ground states with the Pf and APf states.
The degeneracy is indeed lifted between the Pf and the APf, but if one keeps only the lowest Landau level, it is the Pfaffian that is favored. So keeping higher Landau levels when including Landau level mixing is crucial to the get the right physics at small $\kappa$.
At larger $\kappa$, the difference in overlap between the Pf and the APf decreases, and at  $\kappa \approx 1.1$, the ground state of the exact Hamiltonian has a restored particle-hole symmetry, as shown in the inset. 
Now, as could be anticipated from their poor overlap, even-order ground states systematically miss the emerging particle-hole symmetry. By contrast, the odd-order ground states, and in particular the third-order one, agree qualitatively with the restoration of particle-hole symmetry and only underestimate the critical $\kappa_c$.

To summarize, the third-order perturbation theory is in qualitative agreement with the exact result on a small cluster regarding the restoration of particle-hole symmetry, but to include more than the fully occupied Landau level when taking into account Landau level mixing is necessary to get the correct behaviour at small $\kappa$, namely the lifting of the degeneracy in favour of the APf.

\subsection{Fully polarized case}\label{sec:spinless}

In Ref.~\onlinecite{Zaletel2015}, Zaletel \textit{et al} have investigated the ground state of the $\nu = 5/2$ FQHE on an infinite cylinder of finite (but large) circonference using the infinite density matrix renormalization group (iDMRG) algorithm, a method that \emph{exactly} treats Landau level mixing.
In this investigation, the authors of Ref.~\onlinecite{Zaletel2015} make two approximations: they only take into account a small number of Landau levels, and they consider spinless electrons, i.e. they assume full polarization, disregarding completely the filled $0$th Landau level of opposite polarization.
Up to $\kappa = 1.37$, they found that the APf is systematically favored, with no sign of a restoration of the particle-hole symmetry.\\

To compare our perturbative approach with these results, we have performed an investigation of the fully polarized model up to third order in $\kappa$.
Our results do not show any qualitative differences in the ground state properties between $L=4$ and $L=11$, even if the energies vary significantly, so keeping only a few Landau levels, as done in Ref.~\onlinecite{Zaletel2015}, appears to be innocuous. This is not the case of the assumption of full polarization however.
In $H_2$ (resp. $H_3$), the down spins do not contribute to the term with the three-body (resp. four-body) operators due to spin conservation and diagrammatic substraction.
At second-order, only $H_2^\mathrm{3b}$ distinguishes between Pf and APf, and going from spinless to spinful does not change the small $\kappa$ qualitative result.
At third-order, it is no longer the case. The down spins do not change $H_3^\mathrm{4b}$, but they contribute to $H_3^\mathrm{3b}$, and the energy difference induced by this term is reduced for spinless electrons.
As a result, the sum of these two contributions, $H_3$, still favors the Pf, but with a significantly weaker energy difference.
Therefore, the APf regime should a minima survive to larger $\kappa$.

\begin{figure}
\centering\includegraphics[width=\linewidth]{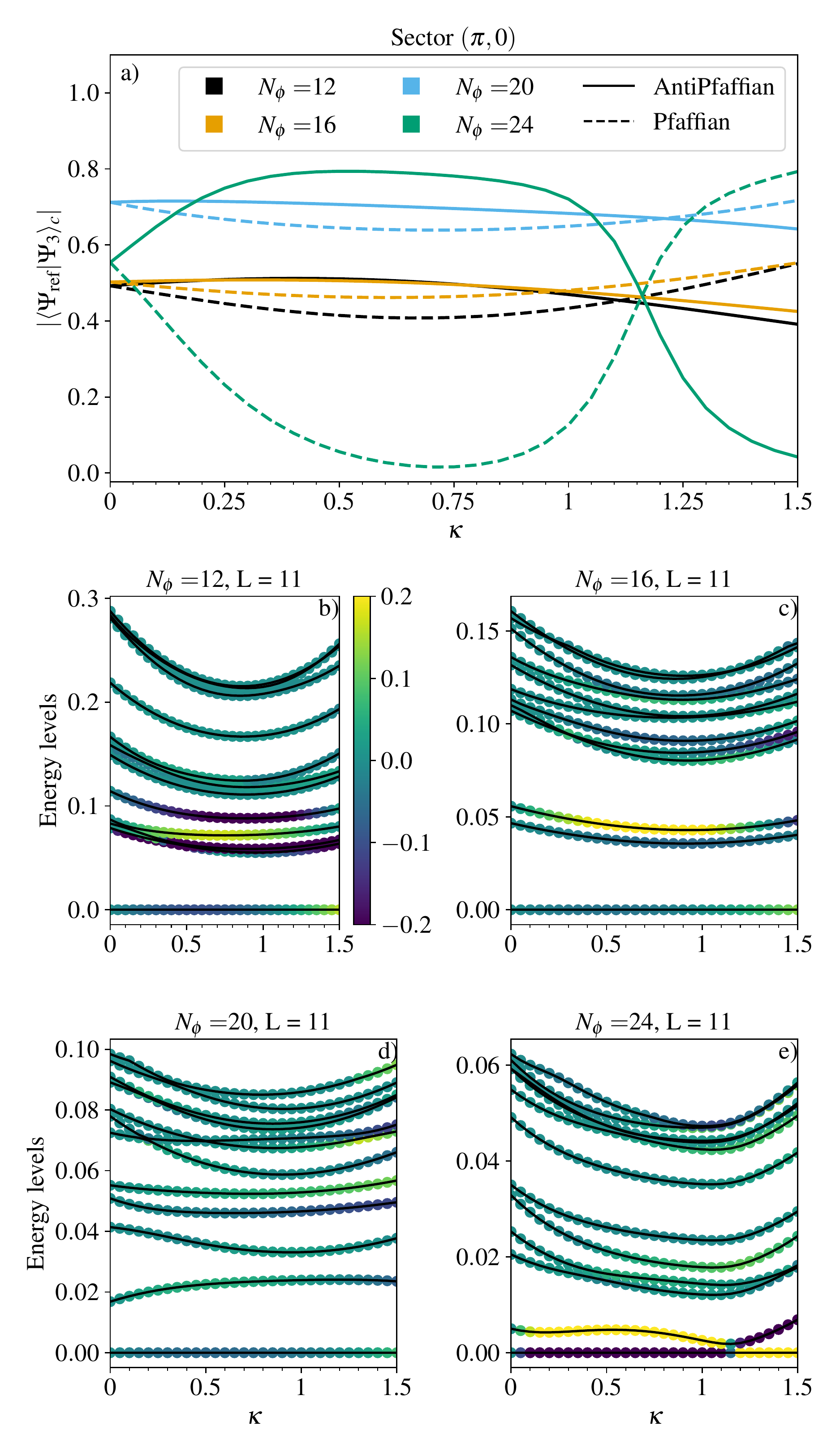}
\caption{(a) Overlaps of the third-order ground state for spinless electrons with the Pf and APf. All overlaps are corrected following Eq.~\eqref{eq:ColorCode}. (b-e) Corresponding low-energy spectra. While the results are qualitatively similar to the spinful case, the critical $\kappa_c \approx 1$ is well beyond the range of the perturbation theory and the energy spectra have less features.}
\label{fig:3rdorderSpinless}
\end{figure}

In order to check this simple argument, we performed both an exact calculation similar to the computation of Sec.~\ref{sec:3exact} for smaller systems, and a third-order expansion at large $N_\phi$.
The results of the perturbative expansion are presented in Fig.~\ref{fig:3rdorderSpinless}.
Overlaps with the Pf and the APf in Fig.~\ref{fig:3rdorderSpinless}\hyperlink{fig:3rdorderSpinless}{a} show a restoration of the particle-hole symmetry at much larger  $\kappa_c \simeq 1.1$, well beyond the range of perturbation theory. We also evaluated the effective Hamiltonian in the Pf-APf subspace as well as their average energies. They also predict a critical $\kappa$ of order $1$.
(see App.~\ref{app:effHam} for details). In order to properly interpret this results, it is important to note that, independent of the exact large-$\kappa$ physics, as long as $H_3$ opposes $H_2$, we will always observe  a crossing of the overlaps at large $\kappa$ at third order. So this result should not be interpreted as evidence that a crossing takes place at a much larger $\kappa_c$. The fact that the crossing occurs for a very large value of $\kappa$ could simply indicate the absence of a restoration of the particle-hole symmetry. \\

We also note that the excitation spectrum shown in Fig.~\ref{fig:3rdorderSpinless}\hyperlink{fig:3rdorderSpinless}{b-e} has significantly less features than in the spinful case, with no obvious tendency to a gap closing, but only an avoided crossing between the two low-lying states for $N_\phi=24$. These results do not suggest the existence of a phase transition in this region.\\

\begin{figure}
\centering\includegraphics[width = 1\linewidth]{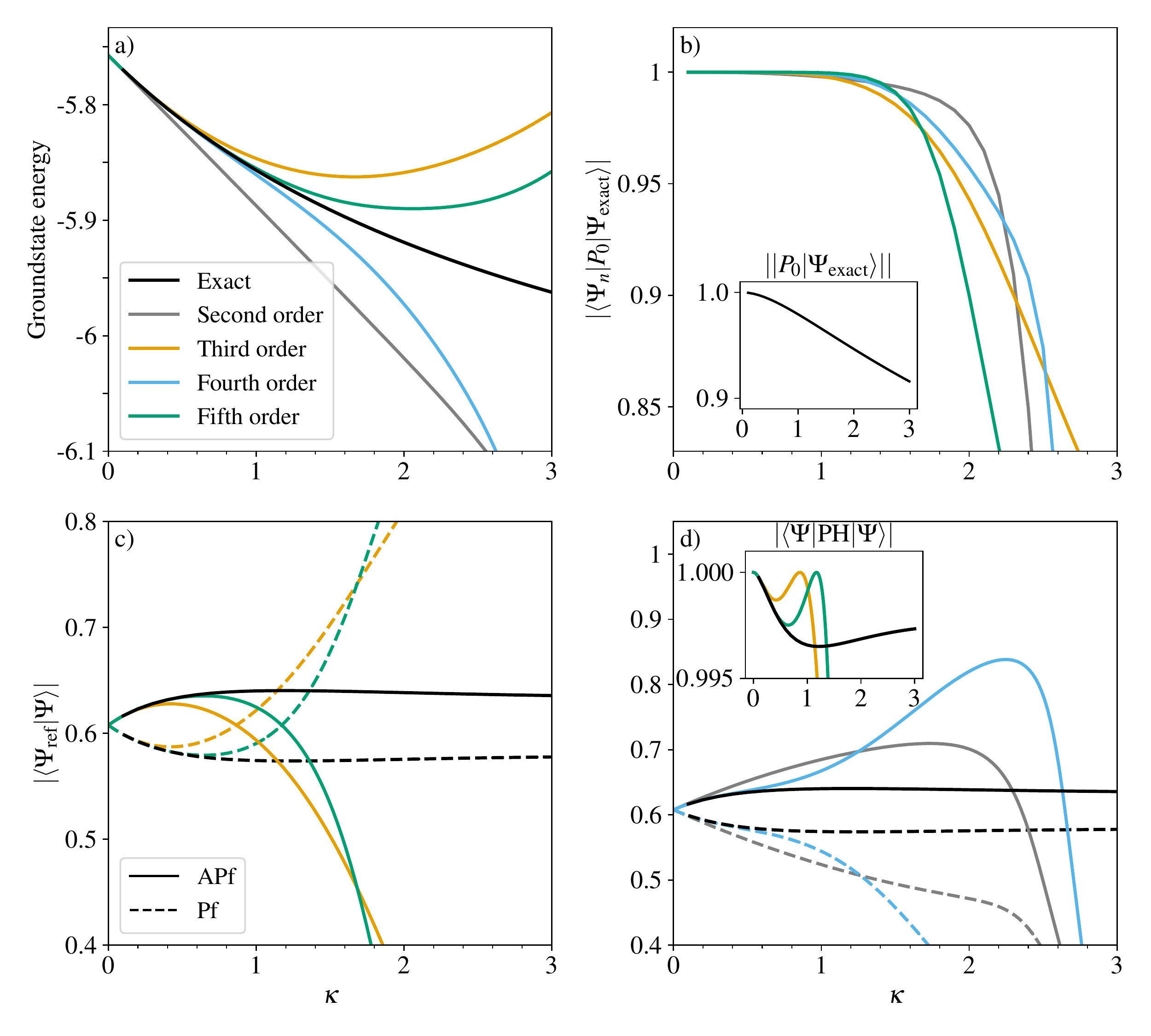}
\caption{a) Energy of the exact ground state for spinless electrons with $L = 2$ and $N_\phi = 8$, compared to the energy obtained from the perturbation theory up to fifth order. b) Overlaps between the perturbative ground states and the normalized projection of the exact ground state. In the inset, we show the norm of the projected ground state. c-d) Comparison of the overlaps with the Pf and the APf for odd (c) and even (d) orders.  The particle-hole symmetry is not restored in the exact ground state, by contrast to all the odd-order perturbative expansions. The splitting remains limited as can be seen from the inset in d) showing the overlaps of the ground state with its particle-hole symmetric partner.}\label{fig:exactspinless}
\end{figure}

The exact calculation presented in Fig.~\ref{fig:exactspinless} confirms this picture.
We computed the ground state of $\mathcal{H}_\mathrm{exact}$ for spinless electrons with $L=3$ and $N_\phi = 8$ and  $N_\phi = 12$.
We only show the results for $N_\phi = 8$ here for comparison with Sec.~\ref{sec:3exact}.
Remarkably enough, there is no restoration of the particle-hole symmetry for the exact ground state.
As expected, all expansions stopping at an odd order predict a crossing in overlaps at sufficiently large $\kappa$,  a clear artifact of truncating the perturbation theory.
The odd orders capture the curvature of the overlaps but overcorrect, which should be expected given that the quasi-saturation we observe can only be visible in the perturbative expansion if the exact ground state at large $\kappa$ is also the ground state of the higher-order term of the series.
Note that the even orders still predict incorrectly the large $\kappa$ limit and severely overestimate the splitting between the Pf and the APf.\\

Altogether, these results point to the possibility of a significant difference between the fully polarized case studied in Ref.~\onlinecite{Zaletel2015} and the case where both spin orientations are allowed for in the lowest Landau level.

\section{Implications for the 5/2 quantum Hall state}

Our numerical results demonstrate unambiguously that, to third-order in $\kappa$, and when the spin of the electrons is taken into account, the particle-hole symmetry is restored in the ground state at $\kappa\simeq 0.4$ regardless of the size of the system. 
What are the physical implications of this result for the actual system? 
The answer is by necessity speculative because we do not have access to higher orders, but two possibilities emerge which could both, to a certain extent, explain the experimental result of the 5/2 thermal conductivity.\\

The first possibility is that the Pf and the APf remain well separated from the other states, and that the physics is simply controlled by their competition. 
The fact that they cross at a fairly small value $\kappa=0.4$ at third order suggests that they will cross at least once even if higher orders are included. This is at the least the case for the $L=2$ model. 
In the vicinity of this crossing the ground state will be almost degenerate, and the explanation of Wang, Vishwanath and Halperin\citep{Wang2018} in terms of Pf and APf domains due to disorder would be viable if the experimental value $\kappa = 1.38$ is close to this level crossing, or to another one if there are several crossings\citep{Hegde2015, Hedge2016, Vionnet2017} as a function of $\kappa$\footnote{This is for instance what happens to the two low-lying states of the Kitaev chain when the pairing is smaller than the hopping, in which case they cross as a function of the chemical potential a number of times equal to the number of sites provided the calculation is exact or at least carried out in perturbation theory to an order equal to the number of sites.}.

The second possibility is that the gap to other excitations closes upon increasing $\kappa$, and that a quantum phase transition occurs at a certain $\kappa_c$ to a phase with an emergent particle-hole symmetry. 
This would lead to a direct explanation of the experimentally measured $5/2$ quantized thermal conductance.
This possibility is supported by the fact that the excited states come down around the point  where particle-hole symmetry is restored at third order before going up again at larger $\kappa$. 
This is typical of the finite-size spectrum of quantum phase transitions as a precursor of criticality in the thermodynamic limit. 
Note that, since the particle-hole symmetry is broken as soon as $\kappa\neq 0$, the symmetry on the other side of the quantum phase transition has to be an emergent one, i.e. the low energy physics is that of a particle-hole symmetric model, but the full spectrum still has traces of the explicit breaking of the particle-hole symmetry in the Hamiltonian. 
If this possibility is realized, it is important to keep in mind that any finite-order calculation, and in particular the third order one reported here, has nothing to say about the nature of the phase beyond the quantum phase transition. 
Indeed, at the quantum phase transition, the energy must be singular, implying that the perturbation in $\kappa$ diverges at that point, or in other words that the expansion in $\kappa$ of the energy has a finite radius of convergence equal to $\kappa_c$. 
So the fact that the Pf is clearly favored for large $\kappa$ must be considered as an artifact of the third order expansion since, for large $\kappa$, the ground state must be that of the higher-order term of the series. 
If the expansion was pushed to higher order, another term would be selected, leading to the stabilization of another state.

In that respect, we would like to note that the first terms of the series contain a hint that the radius of convergence is finite, hence that there is a quantum phase transition.
Indeed, as discussed decades ago in the context of high-temperature expansions of thermal phase transitions\citep{BookDomb}, the radius of convergence is the infinite order limit of the ratio of subsequent coefficients of the series.
The evolution of this ratio with the order can give a hint of the radius of convergence, even if only the first orders are available. 
In particular, if the ratio decreases with the order, this is a strong indication that the radius of convergence is finite, hence that there is a phase transition. 
This is illustrated in Appendix~\ref{app:pertexp} in the case of the 1D transverse field Ising model, for which we show that the ratio decreases smoothly with the order, and that the critical field $h_c=1$ can be accurately deduced from the first three terms of the series by a polynomial fit. 
In the present case, the ratio also {\it decreases} with the order, a behaviour consistent with a finite radius of convergence (see Appendix for details). 
Given the finite order of the series, let us emphasize that this is just an indication and by no means a proof since the argument relies on the assumption that this ratio is monotonous, an assumption that can only be tested by calculating higher orders.\\

To summarize, we believe that our results unambiguously point to a change of behaviour at small field that could explain the puzzling experimental result of a 5/2 thermal conductance. 
This is of course not the end of the story however because the physics at large $\kappa$ is \emph{not} accessible to perturbation theory in $\kappa$. 
To finish, let us briefly discuss the alternative approaches that could possibly shed light on the nature of this state.\\

Since the expansion in $\kappa$ cannot access this phase, one might be tempted to turn to an expansion in $1/\kappa$, starting from small magnetic field. 
Unfortunately, this is probably not going to work either.
At zero magnetic field, the ground state of the 2D interacting electron gas is a Wigner crystal, and there is almost certainly a phase transition between that state and the 5/2 quantum Hall state we are after. 
So this state is probably in an intermediate parameter range that cannot be accessed by perturbation theory from either side.

Turning to variational approaches using wave-functions is also tricky because the wave-functions one can write down explicitly are projected onto a Landau level, and as argued above we do not have access to an accurate effective model in the range of $\kappa$ relevant to experiments.

So it seems that the most promising alternative is to turn to numerics to investigate the intermediate $\kappa$ regime. 
This has already been done using DMRG on cylinders for $\nu=5/2$, but for fully polarized electrons\citep{Zaletel2015}, a case for which the physics at third order is clearly different from the spinful case, as explained in Sec.~\ref{sec:spinless}. 
One obvious suggestion is thus to extend this DMRG calculation to the spinful case. 

Finally, we note that tensor network approaches have been successfully developed for 2D models of quantum many-body physics. 
It would be very interesting to see if these methods can be extended to the FQHE.\\

\acknowledgements
The authors acknowledge useful discussions with Duncan Haldane, Kiryl Pakrouski and Nicolas Regnault. The work has been supported by the Swiss National Science Foundation (FM) grant 182179. Numerical simulations have been performed on the facilities of the Scientific IT and Application Support Center of EPFL.

\bibliographystyle{apsrev4-1}
\bibliography{FQHE}

\appendix

\section{Conventions}\label{app:conventions}
We work with the following form for the Landau level orbitals on the torus:
\begin{multline}
\Psi_{m, l}(x, y) = \frac{1}{\sqrt{L_y\sqrt{\pi} 2^l l! }}
\sum\limits_{k = - \infty}^{\infty} e^{\frac{2 i \pi}{Ly} y (m+kN_\phi)}\\
 H_l(x - \frac{2\pi}{Ly} (m+kN_\phi)) e^{-(x - \frac{2\pi}{Ly} (m+kN_\phi))^2/2}
\end{multline}
We have fixed the gauge such that $\vec{A} = (0, -Bx, 0)$.
Here, we denote by $L_x$ (resp. $L_y$) the dimension of the torus in the $x$ (resp. $y$).
In the main text, we consider a square torus with $L_x = L_y = \sqrt{2 \pi N_\phi}$. 
We checked that this does not affect our result.
We fixed the magnetic length $l_B$ to $1$ for convenience.

\section{Algorithm}\label{app:Algo}
As discussed in the main text, we directly compute the second- and third-order contributions of the empty and occupied Landau levels by using a Schrieffer-Wolff or resolvant type expansion.

The expansion is performed entirely numerically, directly at the operator level in a second-quantization formulation.
It can be divided in several conceptually simple steps which can be easily parallelized.
\begin{enumerate}
\item Computation of the second-quantization coefficients of our chosen interaction for arbitrary Landau levels
\item Generation of all possible Feynmann diagrams
\item Computation of the effective coefficients, taking into account normal-ordering
\item Evaluation of the matrices and determination of their ground states
\end{enumerate}
In the following, we will briefly summarize some of the key points and difficulties of each step.

\subsection{Interaction coefficients}
The computation of second-quantization coefficients is a standard exercice and requires the evaluation of 
\begin{equation}
\langle \Psi_{m_1, l_{m_1}}  \Psi_{m_2, l_{m_2}} \vert V_C \vert  \Psi_{n_1, l_{n_1}}  \Psi_{n_2, l_{n_2}} \rangle
\end{equation}
for all combinations of $m$'s and $l$'s.
The main technical challenge here is the number of such coefficients: $O(N_L^4 N_\phi^3)$.
Translation invariance on the torus allows us to compute only $O(N_L^4 N_\phi^2)$ such terms in practice.
Naively, each coefficient requires a double integration over $\mathbb{R}^2$, which becomes quickly untractable.
Instead, the common approach is to express the interaction in Fourier space
\begin{equation}
V_C = \sum\limits_{\vec{q}} V_C(\vec{q}) e^{i \vec{q}.(\vec{r}_1 - \vec{r}_2)}
\end{equation}
and factorize the computation:
\begin{multline}
\langle \Psi_{m_1, l_{m_1}}  \Psi_{m_2, l_{m_2}} \vert V_C \vert  \Psi_{n_1, l_{n_1}}  \Psi_{n_2, l_{n_2}} \rangle = \\ \sum\limits_{\vec{q}} V_C(\vec{q})  \langle \Psi_{m_1, l_{m_1}} \vert e^{i \vec{q}.\vec{r}} \vert  \Psi_{n_1, l_{n_1}}  \rangle \langle \Psi_{m_2, l_{m_2}} \vert e^{-i \vec{q}.\vec{r}} \vert  \Psi_{n_2, l_{n_2}} \rangle
\end{multline}
Performing the integral on $y$ and after several change of variables, we obtain
\begin{multline}
 \langle \Psi_{m, l_{m}} \vert e^{i \vec{q}.\vec{r}} \vert  \Psi_{n, l_{n}} \rangle = \delta'_{m, n + q_y} \\
  e^{i \frac{q_x q_y}{2}} e^{ i q_x \frac{2 \pi}{L_y} n } e^{-\frac{q^2}{4}} \frac{1}{\sqrt{\pi 2^{l_{m} + l_{n}} l_{m}! l_{n}!}} \\
 \int\limits_{-\infty}^\infty H_{l_{m}}(x -\frac{q_y}{2} + i \frac{q_x}{2} ) H_{l_{n}}(x +\frac{q_y}{2} + i \frac{q_x}{2} ) e^{-x^2}
\end{multline} 
Note that, up to the exponential phase prefactor, this integral no longer depends on $m$ and $n$.
It also decreases exponentially with $q^2$.
We therefore can systematically evaluate it for all values of $q_x$ and $q_y$ up to our precision cutoff, and simply resum the relevant contributions to obtain each of the second quantization coefficients.
In practice, we need of the order of $O(N_\phi^2 L^2)$ such terms for a given precision.
The integral can in fact be computed analytically such that
\begin{multline}
 \langle \Psi_{m, l_{m}} \vert e^{i \vec{q}.\vec{r}} \vert  \Psi_{n, l_{n}} \rangle = \delta'_{m, n + q_y} \\
  e^{i \frac{q_x q_y}{2}} e^{ i q_x \frac{2 \pi}{L_y} n } e^{-\frac{q^2}{4}} \frac{1}{\sqrt{\pi 2^{l_{m} + l_{n}} l_{m}! l_{n}!}} \\
 \sum\limits_{k = 0}^{\min{l_m, l_n}} \binom{l_{m}}{k} \binom{l_{n}}{k} 2^k k! (i q_x - q_y)^{l_m - k} (i q_x + q_y)^{l_n-k}
\end{multline}
We use the above form with heavy memoization in our computation.
Finally, we note that it also accepts an interesting formulation in terms of the Lagrange polynomials:
\begin{multline}
 \langle \Psi_{m, l_{m}} \vert e^{i \vec{q}.\vec{r}} \vert  \Psi_{n, l_{n}} \rangle = \delta'_{m, n + q_y}
  e^{i \frac{q_x q_y}{2}} e^{ i q_x \frac{2 \pi}{L_y} n } e^{-\frac{q^2}{4}} \\
 \left\lbrace\begin{array}{cc}
 L_{l_m}(\frac{q^2}{2})  & \text{if } l_m = l_n\\
 (q_y - i q_x)^{l_m - l_n} \frac{\sqrt{2^{l_n} l_n!}}{\sqrt{2^{l_m} l_m!}} L_{l_m}^{(l_m - l_n)}(\frac{q^2}{2}) &  \text{if } l_m > l_n\\
  (q_y + i q_x)^{l_n - l_m} \frac{\sqrt{2^{l_m} l_m!}}{\sqrt{2^{l_n} l_n!}} L_{l_n}^{(l_n - l_m)}(\frac{q^2}{2}) & \text{if } l_m < l_n
 \end{array}\right.
\end{multline}

\subsection{Feynman diagrams}
The second- and third-order expansions of the Hamiltonian are given in Eq.~\eqref{eq:3rdOrderSW} in the main text.
Terms of the form $P_0 \mathcal{H}_1 \mathcal{G}_0^{m_1} ... \mathcal{H}_1 P_0$ can be naturally expressed using Feynman diagrams.
While we only compute the perturbation up to third-order, we actually have to take into account a large number of diagrams due to the complexity of $\mathcal{H}_1$.
It is convenient to separate the Landau levels in three groups: the fully occupied levels, the fully empty ones, and our target.
Each operator $c$ or $c^\dag$ then belongs to $3$ possible groups.
For convenience, we strictly order the $c$ (resp. $c^\dag$) by the label of their Landau levels (and by spin in case of equality).
$\mathcal{H}_1$ is then separated into $36$ terms, each corresponding to a possible vertex in the Feynman representation of the perturbative expansion.
At second-order, it is still possible to implement by hand the relevant diagrams: $27$ diagrams to obtain all contributions, and only $12$ for the non-diagonal ones (neglecting all symmetries).
At third-order, we have to deal with several hundreds diagrams ($835$ diagrams if we want all contributions, and $458$ if we are only interested in the non-diagonal contributions)\footnote{In practice, this number can be reduced by a factor $2$ by a combination of Hermitian conjugation and anticommutation} and it is necessary to do that automatically.
Straightforward linear programming with contraints easily lists all the relevant combination of vortices, which we use as inputs for the last block of our code. \\

Note that we compute Feynman diagrams leading to diagonal terms in our expansion as they are necessary for the computation of the anticommutators appearing in the third-order contributions.
Their computational cost is negligible.

\subsection{Effective interactions}
Once we have the list of diagrams, we can compute the effective Hamiltonian in second-quantized form.
To do so, we automatically generate all possible combinations of Landau levels and orbitals allowed by the vertices.
This can again be done using programmation under constraints and memoization to speed up and limit redudant evaluations.
The number of orbitals in a given diagram scales approximately as $O(N_\phi^{n_t + n_o/2 + n_e/2 - 3} L_o^{n_0}  L_e^{n_e})$,
where $n_{e/o/t}$ is the number of operators in empty (resp. occupied and targeted) Landau levels, and $L_{e/o}$ the number of empty (resp. occupied) Landau levels.
Taking into account the translation invariance allows us to reduce the complexity by an additional factor $N_\phi$ at the price of symmetrization.\\

The last step is to normal-order the obtained interactions to keep the number of coefficients to implement manageable.
As the complexity is exponential, it is key to implement the normal ordering in a memory-allocation and complexity optimal fashion.
The most expensive diagrams to compute are those leading to the five-body terms.
They contribute to lower-body terms after normal-ordering, even though the five-body interaction itself vanishes as the contribution from $P_0 \mathcal{H}_1 \mathcal{G}_0  \mathcal{H}_1 \mathcal{G}_0  \mathcal{H}_1 P_0$ and the anticommutator exactly cancels.\\

Parallelization can be realized either at the diagram levels, or within the diagrams themselves.
We chose the latter for convenience (at the price of a larger memory footprint).
For the largest system sizes, it is also more efficient to compute the anticommutator in Eq.~\eqref{eq:3rdOrderSW} directly at the operator levels.

\subsection{Evaluation of the Hamiltonian}

Finally, the evaluation of the Hamiltonian from the second-quantization coefficients is straighforward.
The main limitation lies in the density of the Hamiltonians which restrict the achievable system sizes.
For the largest size considered here $N_\phi = 28$, more than $10$ millions operators are present in the Hamiltonian if we discard all contributions smaller than $10^{-10}$ (and $3$ millions if we choose a cutoff at $10^{-6}$).
This translates into Hamiltonians that have $25\%$ filling even for $N_\phi = 28$, strongly limiting our possibilities.
$N_\phi = 32$, the next relevant size for paired systems (such that the number of electrons remains even), admits $\approx 10^6$ vectors in the translation-invariant basis.
Even a $10\%$ filled matrix would occupy approximately $1TB$ of RAM (using 64-bit floating points), which is technically achievable, but numerically heavy to build and use.
Similarly, applying the Hamiltonian on the fly without constructing it would require several thousand of billions of operations per vector application.
While technically achievable, we limited ourselves to $N_\phi = 28$ given the massive numerical costs.\\

The explosion of the number of coefficients will actually also hard-limit the possible orders of the expansion as long as we keep this second-quantized formulation.
Even including only the five-body terms appearing at the next order would require the computation and evaluation of $100$ millions operators per basis state.
Given the results of the exact calculation for small systems (even-order expansions are physically wrong faster than the odd-order), it appears impractical or impossible to extend this naive expansion up to fifth-order in the foreseeable future.
Other approaches such as DMRG would be more appropriate.

\section{Effective Hamiltonians between Pfaffian and AntiPfaffian}\label{app:effHam}
This section summarizes the effective Hamiltonian we obtained in the Pf-APf subspace.
They are included in this Apppendix both as a reference for benchmarking, and to illustrate the opposite effects of the second and third-order perturbative expansion.
Concretely, we define the reduced effective Hamiltonian
\begin{equation}
h_n = \frac{1}{N_{l=1}}\begin{pmatrix}
\Psi_\mathrm{Pf} & \Psi_\mathrm{APf} 
\end{pmatrix} H_n \begin{pmatrix}
\Psi_\mathrm{Pf} \\ \Psi_\mathrm{APf} 
\end{pmatrix}. \label{eq:projHam}
\end{equation}
Due to the non-orthogonality of the states, we also investigate the corrected effective Hamiltonian
\begin{equation}
\tilde{h}_n = M^{-\frac{1}{2}} h_n M^{-\frac{1}{2}}
\end{equation}
with $M$ defined in Eq.~\eqref{eq:ColorCode}.
$\tilde{h}_n$ is in principle a better measure of the energy splitting and the coupling between Pf and APf.
Our conclusions are nonetheless unaffected.\\

First, we show the results for the spinful system.
In Tabs.~\ref{tab:effham}, we directly list the effective Hamiltonian $h_n$ defined in Eq.~\eqref{eq:projHam} in the Pauli matrix basis.
In Tab.~\ref{tab:effhamcorr}, we present $\tilde{h}_n$ for the same systems.
$h_n$ sees more significant finite-size effects, even at fixed $L$.
The crossing point is reliably at $\kappa_c \approx 0.4$.
The Padé $(1, 1)$ approximant of the series $(h_1, h_2, h_3)$ gives similar results, with a restoration of the particle-hole symmetry at $\kappa_c \approx 0.6$.
In Tabs.~\ref{tab:effhamspinless} and \ref{tab:effhamcorrspinless}, we show the same quantities for spinless systems, with $\kappa_c \approx 1$.
The influence of the third-order expansion is reduced.
If we perform the same Padé $(1, 1)$ approximant, $\kappa_c$ is pushed above $2$ with significant variations with the system size.
Note nonetheless that this approximant is here extremely sensitive to the details of the computation: Were we to not include the trivial constant terms in $H_3$, we would see no crossing for any $N_\phi$.

\begin{table*}
\begin{tabular}{c|c|c|c|c|c}
$N_\phi$ & Sector & $L$ & $h_1$ & $h_2$ & $h_3$ \\
$ 28 $ & $ (\pi, 0) $ & $ 4 $ & $ (-1.3829, 0, -0.2279) $ & $ (-0.1983, 0.0004, -0.0322) $ & $ (0.0657, -0.0009, 0.0103) $ \\
$ 28 $ & $ (\pi, \pi) $ & $ 4 $ & $ (-1.3828, 0, -0.6516) $ & $ (-0.1983, 0.0003, -0.0925) $ & $ (0.0652, -0.0008, 0.0291) $ \\
$ 24 $ & $ (\pi, 0) $ & $ 11 $ & $ (-1.354, 0, -0.3511) $ & $ (-0.3087, 0.0004, -0.0796) $ & $ (0.1024, -0.0009, 0.0256) $ \\
$ 24 $ & $ (\pi, \pi) $ & $ 11 $ & $ (-1.3539, 0, -0.4565) $ & $ (-0.3087, 0.0004, -0.1038) $ & $ (0.102, -0.001, 0.0334) $\\
$ 20 $ & $ (\pi, 0) $ & $ 11 $ & $ (-1.3172, 0, -0.4494) $ & $ (-0.3084, 0.0003, -0.1049) $ & $ (0.102, -0.0008, 0.0338) $ \\
$ 20 $ & $ (\pi, \pi) $ & $ 11 $ & $ (-1.3172, 0, -0.118) $ & $ (-0.3085, 0.0004, -0.0274) $ & $ (0.1024, -0.0009, 0.0086) $ \\
$ 16 $ & $ (\pi, 0) $ & $ 11 $ & $ (-1.268, 0, -0.658) $ & $ (-0.308, 0.0003, -0.1581) $ & $ (0.1024, -0.0008, 0.0514) $ \\
$ 16 $ & $ (\pi, \pi) $ & $ 11 $ & $ (-1.2687, 0, -0.9211) $ & $ (-0.3074, 0.0003, -0.2207) $ & $ (0.1029, -0.0007, 0.0725) $ \\
$ 12 $ & $ (\pi, 0) $ & $ 11 $ & $ (-1.1948, 0, -0.9699) $ & $ (-0.3088, 0.0005, -0.2488) $ & $ (0.1028, -0.0011, 0.0812) $ \\
$ 12 $ & $ (\pi, \pi) $ & $ 11 $ & $ (-1.1947, 0, -0.4337) $ & $ (-0.3104, 0.0005, -0.1107) $ & $ (0.1077, -0.0013, 0.0382) $
\end{tabular}
\caption{Effective Hamiltonians in the Pf-APf subspace for spinful electrons, defined in Eq.~\eqref{eq:projHam}. We use the notation $h_n = n_0  \sigma^0 + n_x \sigma^x + n_z \sigma^z \equiv (n_0, n_z, n_x)$.}
\label{tab:effham}
\end{table*}

\begin{table*}
\begin{tabular}{c|c|c|c|c|c}
$N_\phi$ & Sector & $L$ & $\tilde{h}$ & $\tilde{h}_2$ & $\tilde{h}_3$ \\
$ 28 $ & $ (\pi, 0) $ & $ 4 $ & $ (-1.3828, 0, -0.0003) $ & $ (-0.1984, 0.0004, 0.0005) $ & $ (0.0658, -0.0009, -0.0005) $ \\
$ 28 $ & $ (\pi, \pi) $ & $ 4 $ & $ (-1.3826, 0, -0.0006) $ & $ (-0.1988, 0.0004, 0.0011) $ & $ (0.0661, -0.0009, -0.002) $ \\
$ 24 $ & $ (\pi, 0) $ & $ 11 $ & $ (-1.354, 0, -0.0001) $ & $ (-0.3088, 0.0004, 0.0004) $ & $ (0.1027, -0.0009, -0.001) $ \\
$ 24 $ & $ (\pi, \pi) $ & $ 11 $ & $ (-1.3538, 0, -0.0001) $ & $ (-0.3088, 0.0004, 0.0004) $ & $ (0.1024, -0.001, -0.0011) $\\
$ 20 $ & $ (\pi, 0) $ & $ 11 $ & $ (-1.3172, 0, -0.0001) $ & $ (-0.3085, 0.0004, 0.0004) $ & $ (0.1024, -0.0008, -0.0012) $\\
$ 20 $ & $ (\pi, \pi) $ & $ 11 $ & $ (-1.3172, 0, 0) $ & $ (-0.3085, 0.0004, 0.0002) $ & $ (0.1025, -0.0009, -0.0006) $\\
$ 16 $ & $ (\pi, 0) $ & $ 11 $ & $ (-1.2671, 0, -0.0016) $ & $ (-0.309, 0.0004, 0.002) $ & $ (0.1036, -0.0009, -0.0023) $\\
$ 16 $ & $ (\pi, \pi) $ & $ 11 $ & $ (-1.2659, 0, -0.0038) $ & $ (-0.3107, 0.0004, 0.0045) $ & $ (0.1059, -0.001, -0.0042) $\\
$ 12 $ & $ (\pi, 0) $ & $ 11 $ & $ (-1.1921, 0, -0.0033) $ & $ (-0.3125, 0.0008, 0.0046) $ & $ (0.1079, -0.0019, -0.0063) $\\
$ 12 $ & $ (\pi, \pi) $ & $ 11 $ & $ (-1.194, 0.0, -0.0019) $ & $ (-0.3111, 0.0006, 0.0018) $ & $ (0.108, -0.0014, -0.0009) $
\end{tabular}
\caption{Effective Hamiltonians $\tilde{h}_n = M^{-\frac{1}{2}} h_n M^{-\frac{1}{2}}$ in the Pf-APf subspace corrected by the overlap matrix for spinful electrons. We round at the fourth decimal.}
\label{tab:effhamcorr}
\end{table*}

\begin{table*}
\begin{tabular}{c|c|c|c|c|c}
$N_\phi$ & Sector & $L$ & $h_1$ & $h_2$ & $h_3$ \\
$ 24 $ & $ (\pi, 0) $ & $ 11 $ & $ (-0.8849, 0, -0.2295) $ & $ (-0.0289, 0.0004, -0.0071) $ & $ (0.0067, -0.0003, 0.0015) $\\
$ 24 $ & $ (\pi, \pi) $ & $ 11 $ & $ (-0.8848, 0, -0.2984) $ & $ (-0.0287, 0.0004, -0.0092) $ & $ (0.0066, -0.0003, 0.002) $ \\
$ 20 $ & $ (\pi, 0) $ & $ 11 $ & $ (-0.8629, 0, -0.2944) $ & $ (-0.0288, 0.0003, -0.0093) $ & $ (0.0066, -0.0003, 0.002) $ \\
$ 20 $ & $ (\pi, \pi) $ & $ 11 $ & $ (-0.8629, 0, -0.0773) $ & $ (-0.029, 0.0004, -0.0024) $ & $ (0.0067, -0.0003, 0.0005) $\\
$ 16 $ & $ (\pi, 0) $ & $ 11 $ & $ (-0.8335, 0, -0.433) $ & $ (-0.029, 0.0003, -0.0144) $ & $ (0.0069, -0.0003, 0.0032) $\\
$ 16 $ & $ (\pi, \pi) $ & $ 11 $ & $ (-0.8342, 0, -0.6063) $ & $ (-0.0293, 0.0003, -0.0206) $ & $ (0.0071, -0.0003, 0.0048) $\\
$ 12 $ & $ (\pi, 0) $ & $ 11 $ & $ (-0.7892, 0, -0.641) $ & $ (-0.0286, 0.0005, -0.0224) $ & $ (0.0068, -0.0004, 0.005) $\\
$ 12 $ & $ (\pi, \pi) $ & $ 11 $ & $ (-0.7891, 0, -0.287) $ & $ (-0.0306, 0.0005, -0.0108) $ & $ (0.0077, -0.0005, 0.0025) $
\end{tabular}
\caption{Effective Hamiltonians in the Pf-APf subspace for spinless electrons, defined in Eq.~\eqref{eq:projHam}. We use the notation $h_n = n_0  \sigma^0 + n_x \sigma^x + n_z \sigma^z \equiv (n_0, n_z, n_x)$.}
\label{tab:effhamspinless}
\end{table*}

\begin{table*}
\begin{tabular}{c|c|c|c|c|c}
$N_\phi$ & Sector & $L$ & $\tilde{h}$ & $\tilde{h}_2$ & $\tilde{h}_3$ \\
$ 24 $ & $ (\pi, 0) $ & $ 11 $ & $ (-0.8849, 0, -0.0001) $ & $ (-0.0291, 0.0004, 0.0004) $ & $ (0.0067, -0.0003, -0.0002) $\\
$ 24 $ & $ (\pi, \pi) $ & $ 11 $ & $ (-0.8848, 0, -0.0001) $ & $ (-0.0289, 0.0004, 0.0005) $ & $ (0.0066, -0.0003, -0.0003) $\\
$ 20 $ & $ (\pi, 0) $ & $ 11 $ & $ (-0.8629, 0, -0.0001) $ & $ (-0.029, 0.0004, 0.0005) $ & $ (0.0067, -0.0003, -0.0003) $\\
$ 20 $ & $ (\pi, \pi) $ & $ 11 $ & $ (-0.8629, 0, 0) $ & $ (-0.029, 0.0004, 0.0002) $ & $ (0.0067, -0.0003, -0.0001) $\\
$ 16 $ & $ (\pi, 0) $ & $ 11 $ & $ (-0.8327, 0, -0.0016) $ & $ (-0.0294, 0.0004, 0.0008) $ & $ (0.0071, -0.0003, -0.0005) $\\
$ 16 $ & $ (\pi, \pi) $ & $ 11 $ & $ (-0.8315, 0, -0.0038) $ & $ (-0.0303, 0.0004, 0.0014) $ & $ (0.0076, -0.0004, -0.0008) $\\
$ 12 $ & $ (\pi, 0) $ & $ 11 $ & $ (-0.7866, 0, -0.0033) $ & $ (-0.0305, 0.0008, 0.0023) $ & $ (0.0079, -0.0006, -0.0014) $\\
$ 12 $ & $ (\pi, \pi) $ & $ 11 $ & $ (-0.7884, 0, -0.0019) $ & $ (-0.0307, 0.0006, 0.0003) $ & $ (0.0078, -0.0005, -0.0003) $
\end{tabular}
\caption{Effective Hamiltonians $\tilde{h}_n = M^{-\frac{1}{2}} h_n M^{-\frac{1}{2}}$ in the Pf-APf subspace corrected by the overlap matrix for spinless electrons.}
\label{tab:effhamcorrspinless}
\end{table*}

\section{Perturbative expansion and quantum phase transition}\label{app:pertexp}

In this Appendix, we discuss in more details the connection between the perturbative expansion of the ground state energy of a Hamiltonian as a function of a parameter, say $x$, and the occurence of a quantum phase transition at a critical parameter $x_c$. In the context of thermal phase transitions, this has been discussed at length\citep{BookDomb} for high-temperature expansions. If a phase transition occurs at inverse temperature $\beta_c$, this means that the free-energy has a singularity, which in turns implies that the radius of convergence of the series must be equal to $\beta_c$. By analogy, a quantum phase transition can be expected to take place at $x_c$ if the radius of convergence of the series of the ground-state energy is finite and equal to $x_c$.

Now, if we denote by $a_n$ the coefficients of the expansion of the ground-energy, the radius of convergence is given by the limit of $|a_{n-1}/a_{n}|$ when $n$ goes to infinity. In principle, this implies that one should know the series to infinite order to know if there is a quantum critical point. However, if the series is well behaved, the first terms of the series can already contain some relevant information. In particular, if the ratio $|a_{n-1}/a_{n}|$ decreases with $n$, the radius of convergence cannot be infinite, and a quantum phase transition has to take place.

Let us consider as an example the transverse field Ising model whose Hamiltonian is
\begin{equation}
H = - J \sum\limits_j \sigma^x_j \sigma^x_{j+1} - h \sum\limits_j \sigma^z_j.
\end{equation}
The exact expression for the ground state energy as a function of the transverse field $h$\citep{BookChakrabarti} can be used to evaluate the coefficients of the expansion. All odd coefficients vanish, and the even coefficients are given by $a_0=-1$ and,
\begin{equation}
a_n=-\prod_{p=2,p\  \text{even}}^{n}\left(\frac{p-3}{p}\right)^2, n\geq 2
\end{equation}
For $n\geq 4$, $(a_{n-2}/a_{n})^{1/2}=\frac{n}{n-3}$. The radius of convergence is thus given by 
\begin{equation}
h_c=\lim_{n\rightarrow \infty} (a_{n-2}/a_{n})^{1/2}=\lim_{n\rightarrow \infty} 1+\frac{3}{n-3}=1
\end{equation}
or equivalently by
\begin{equation}
1/h_c=\lim_{n\rightarrow \infty} (a_{n}/a_{n-2})^{1/2}=\lim_{n\rightarrow \infty} 1-\frac{3}{n}=1
\end{equation}
The radius of convergence is $h_c=1$, in agreement with the exact result, but more importantly the ratio $(a_{n-2}/a_{n})^{1/2}$ decreases with $n$ for $n\geq 4$. Besides, expressed as a function of $1/n$, the ratio giving $1/h_c$, $(a_n/a_{n-2})^{1/2}$, reduces to a polynomial of degree 1. Hence, a linear regression of this ratio for the first two terms $n=4$ and $n=6$ gives the correct limit, $h_c=1$.

For our model, we have computed the derivatives of the ground state energy for the third-order perturbation Hamiltonian, and for the exact Hamiltonian with $L = 2$. They are listed in Tab.~\ref{tab:perturbativeenergies}.
In terms of the derivatives, coefficients and their ratio are given by:
\begin{align}
a_n &= \frac{1}{n!} \partial_\kappa^n E(\kappa = 0).\\
a_{n-1}/a_n &= n \partial_\kappa^{n-1} E(\kappa = 0)/ \partial_\kappa^n E(\kappa = 0).
\end{align}
For our ground state, the ratio $a_{n-1}/a_n$ also decreases between $n=1$ and $n=2$.
Taken naively at $n=2$, it gives $\kappa_c \approx 3 $ for $L = 4$ and $L = 10$.
These values of course are but naive approximations, and should just be seen as a suggestion of the existence of a critical point.
For $L=2$, we extracted the first four derivative following our exact computation, and we found $\kappa_c \approx 2$ performing a naive linear regression.

\begin{table*}[h!]
\begin{tabular}{c|c|c|c|c|c|c|c}
$N_\phi$ & $L$ & $ E_3(\pi, 0)$ & $ \partial_\kappa E_3(\pi, 0)$ & $ \partial^2_\kappa E_3(\pi, 0)$ & $ E_3(\pi, \pi)$ & $ \partial_\kappa E_3(\pi, \pi)$ & $ \partial^2_\kappa E_3(\pi, \pi)$ \\
$ 28 $ & $ 4 $ & $  -1.3839 $ & $ -0.1974 $ & $ 0.1316 $ & $ -1.384 $ & $ -0.1973 $ & $ 0.1322 $ \\
$ 24 $ & $ 11 $ & $  -1.3548 $ & $ -0.3081 $ & $ 0.2044 $ & $ -1.3549 $ & $ -0.3085 $ & $ 0.2063 $ \\
$ 24 $ & $ 4 $ & $  -1.3548 $ & $ -0.1979 $ & $ 0.1316 $ & $ -1.3549 $ & $ -0.1982 $ & $ 0.134 $ \\
$ 20 $ & $ 11 $ & $  -1.3187 $ & $ -0.308 $ & $ 0.2087 $ & $ -1.3181 $ & $ -0.3082 $ & $ 0.2077 $ \\
$ 20 $ & $ 4 $ & $  -1.3187 $ & $ -0.1976 $ & $ 0.1365 $ & $ -1.3181 $ & $ -0.1979 $ & $ 0.1353 $ \\
$ 16 $ & $ 11 $ & $  -1.27 $ & $ -0.3062 $ & $ 0.2045 $ & $ -1.2704 $ & $ -0.3059 $ & $ 0.2044 $ \\
$ 16 $ & $ 4 $ & $  -1.27 $ & $ -0.1959 $ & $ 0.1322 $ & $ -1.2704 $ & $ -0.1955 $ & $ 0.1323 $ \\
$ 12 $ & $ 11 $ & $  -1.1963 $ & $ -0.3076 $ & $ 0.2055 $ & $ -1.1959 $ & $ -0.3093 $ & $ 0.2142 $ \\
$ 12 $ & $ 4 $ & $  -1.1963 $ & $ -0.1973 $ & $ 0.1326 $ & $ -1.1959 $ & $ -0.199 $ & $ 0.142 $\\
$N_\phi$ & $L$ & $ E(\pi, 0)$ & $ \partial_\kappa E(\pi, 0)$ & $ \partial^2_\kappa E(\pi, 0)$ & $ \partial^3_\kappa E(\pi, 0)$ & $ \partial^4_\kappa E(\pi, 0)$  \\
$8$ & $2$ & $-1.0778$ &  $-0.04912$ & $ 0.04228$ & $-0.0590$   & $0.115$ 
\end{tabular}
\caption{ground state energy per particle and its first two derivatives estimated from the third-order perturbative expansion for different $N_\phi$ and $L$. Higher derivative cannot be estimated from the expansion. For $N_\phi = 8$, we performed the computation exactly, and we have access to higher-orders of the derivative.}
\label{tab:perturbativeenergies}
\end{table*}

\end{document}